%% file: main.tex
\documentclass[final,twocolumn]{elsarticle}
\usepackage[switch]{lineno}
\usepackage{hyperref}
\usepackage{color,soul}
\usepackage{amsmath,amssymb,amsfonts}
\usepackage{array}
\usepackage{multirow}
\usepackage[table]{xcolor}
\usepackage{xcolor}
\usepackage[linesnumbered,ruled,vlined]{algorithm2e}
\usepackage{epstopdf}
\usepackage{algorithmic}
\usepackage{graphicx}
\usepackage{textcomp}
\usepackage{caption}
\usepackage{subcaption}
\usepackage{multido}
\usepackage{booktabs}
\usepackage{array}

\makeatletter
\newcommand{\pushright}[1]{\ifmeasuring@#1\else\omit\hfill$\displaystyle#1$\fi\ignorespaces}
\newcommand{\pushleft}[1]{\ifmeasuring@#1\else\omit$\displaystyle#1$\hfill\fi\ignorespaces}
\makeatother
\makeatletter
\def\hlinewd#1{%
	\noalign{\ifnum0=`}\fi\hrule \@height #1 \futurelet
	\reserved@a\@xhline}
\makeatother

\input{new_comms.tex}
\begin{document}
\begin{frontmatter}

\title{Twin-in-the-loop state estimation for vehicle\\dynamics control: theory and experiments}

\author[first]{Giorgio Riva}
\author[first]{Simone Formentin\corref{mycorrespondingauthor}}\ead{simone.formentin@polimi.it}
\author[first]{Matteo Corno} 
\author[first]{Sergio M. Savaresi}

\address[first]{Dipartimento di Elettronica, Informazione e Bioingegneria, Politecnico di Milano, P.za L. da Vinci 32, 20133 Milano, Italy\\}
\cortext[mycorrespondingauthor]{Corresponding author.}
\begin{abstract}
In vehicle dynamics control, many variables of interest cannot be directly measured, as sensors might be costly, fragile or even not available. Therefore, real-time estimation techniques need to be used. The previous approach suffers from two main drawbacks: (i) the approximations due to model mismatch might jeopardize the performance of the final estimation-based control; (ii) each new estimator requires the calibration from scratch of a dedicated model.
In this paper, we propose a \textit{twin-in-the-loop} scheme, where the ad-hoc model is replaced by an accurate multibody simulator of the vehicle, typically available to vehicles manufacturers and suitable for the estimation of any on-board variable, coupled with a compensator within a closed-loop observer scheme. Given the black-box nature of the digital twin, a data-driven methodology for observer tuning is developed, based on Bayesian optimization. 
The effectiveness of the proposed estimation method for the estimation of vehicle states and forces, as compared to traditional model-based Kalman filtering, is experimentally shown on a dataset collected with a sport car. 
\end{abstract}

\begin{keyword}
State estimation; Observer design; Vehicle dynamics
\end{keyword}

\end{frontmatter}
\section{Introduction}
\label{sec:introduction}
Vehicle dynamics control represents a cornerstone in automotive to enhance performance and safety on commercial vehicles, e.g. in Anti-lock Braking Systems (ABS) and stability control. 
Indeed, due to economic and physical constraints, many variables needed by vehicle dynamics controllers cannot be directly measured, calling for estimation solutions that exploit the available measurements and suitable dedicated models of selected vehicle dynamics, see \cite{singh2018literature}, \cite{viehweger2020vehicle}.

\textbf{Analysis of the state of the art.}
The vehicle state estimation literature is quite broad, spanning among kinematic and dynamic variables, and vehicle parameters. Longitudinal velocity, side-slip angles, and tire-road forces are the most common examples in the former class, while mass, inertia and the maximum friction coefficient are the typical estimated parameters. A comprehensive review of the state-of-the-art solutions is available in \cite{singh2018literature}. The estimation of the vehicle longitudinal speed, which is fundamental to control the longitudinal slip during braking manoeuvers \cite{savaresi_braking}, mainly relies on kinematic models. For instance, the authors of \cite{tanelli2006longitudinal} propose an algorithm based only on wheel velocities, the method in \cite{WANG_2004} employs a moving-horizon strategy based on a simple acceleration model, while the three-axis velocities are jointly estimated by \cite{oh2011design} using an Adaptive Kalman Filter (AKF).
Vehicle side-slip angle estimation, instead, is addressed through different approaches, from kinematic-based, like in \cite{selmanaj2017vehicle,breschi2020vehicle}, to dynamic-based ones \cite{madhusudhanan2016vehicle}, which exploit both single-track and double-track models, \cite{DOUMIATI_ACC2010,DOUMIATI_IEEE_TM}.
The objective of vehicle dynamics estimators is broader than planar kinematic variables. Typical parameters of interest are vehicle inertias: in \cite{FATHY_2008} vehicle mass is estimated via a Recursive Least Square (RLS) algorithm, \cite{isgro2018line} proposes a classification approach for passenger detection in motorcycles, in \cite{VAHIDI_2005} mass and road slope are jointly estimated, while \cite{rozyn2010method} proposes a method to retrieve the whole set of inertia parameters. Another important parameter is the maximum friction, crucial for determining the achievable level of safety and performance, see \cite{acosta2017road} and \cite{selmanaj2019friction}.

Apart from vehicle-related parameters, and linked to the maximum friction coefficient, also the knowledge of tire/road forces is of paramount importance because the behaviour of the vehicle is ultimately determined by such variables, see, e.g., \cite{corno2011hybrid}, \cite{madhusudhanan2013lateral}, \cite{jiang2019real}. However, the most accurate measurement device is represented by dynamo metric wheels \cite{BLASCO_2014}, whose diffusion is limited by economic and space requirements. Other interesting solutions under development are represented by load sensing bearings \cite{van2005measurement},\cite{madhusudhanan2016load} and in-tire sensors \cite{KRIER_2014}, which are not ready for commercial production. As a consequence, the problem of estimating tire forces has gained a lot of attention in the last decades. Planar forces, namely longitudinal and lateral ones, are typically estimated through single-track and double-track models. For instance, \cite{BAFFET20091255} and \cite{ZhangICEMI} propose a Sliding Mode Observer (SMO) to jointly estimate axle forces, \cite{WILKIN} uses a similar approach employing an Extended Kalman Filter (EKF), extended by \cite{HAMANN_IVS} with an Unscented Kalman Filter (UKF) approach. \cite{MSIRDI_VSD} enlarges the single track model with the relaxation dynamics in a SMO approach. In \cite{rezaeian2014novel}, lateral forces are estimated making use of longitudinal forces, which are obtained from a wheel dynamics model, as done also in  \cite{HSIAO_2011}, which introduces the friction ellipse in the estimation process. About the double-track model, \cite{DOUMIATI_ACC2010} proposes an EKF approach, while \cite{DOUMIATI_IEEE_TM,DOUMIATI_ECC2009} discuss a comparison between EKF and UKF about lateral forces estimation. A SMO application is reported by \cite{SHRAIM_2007}, while \cite{DOUMIATI_ACC2010} proposes a force split at each wheel based on estimated vertical forces. Finally, concerning the estimation of vertical forces, \cite{DOUMIATI_2008_IFAC} employs a linear quarter-car model based on a Kalman Filter (KF) approach, while \cite{JIANG_2015} couples the four corners by exploiting pitch and roll dynamics.

\textbf{Major limitations.} Although different, all the above mentioned works share one common feature: an ad-hoc model (describing the dynamics of interest) is proposed for each variable to estimate, so that many different estimation modules are required and must be calibrated separately. Moreover, concerning tire-road forces, the estimators do not provide per-wheel estimation results, and simple heuristics for force allocation need to be introduced to split the estimated forces. Also this fact is due to the employed simplified models, which, although simple to tune and implement, provide limited description capabilities and accuracy.

\begin{figure}[h!]
	\centering
	\includegraphics[width=\columnwidth]{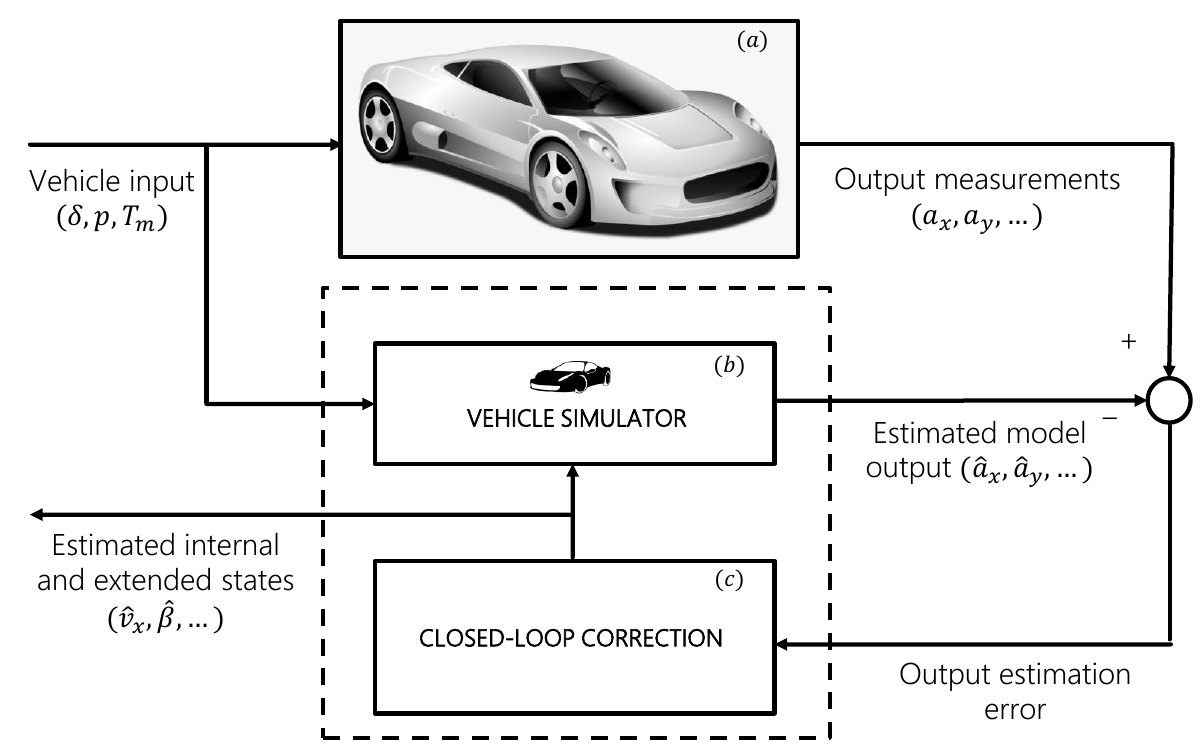}
	\caption{Twin-in-the-loop estimation framework: the synergy between the vehicle digital twin and the closed-loop correction modules allows to estimate vehicle states  employing available online measurements.}
	\label{fig:est_problem}
\end{figure}

\textbf{Novel contributions.} This paper aims to provide a \textit{unified estimator for all the variables of interest in vehicle dynamics}, e.g. longitudinal and lateral vehicle velocities, parameters and tire road forces. The approach is grounded on the classical closed-loop estimation approach, typical of many solutions proposed in the literature, whereas the main innovation is twofold: (1) the use of a simulation-oriented multibody vehicle model, in contrast to control-oriented ones, and (2) the subsequent data-driven calibration procedure, because the simulator black-box nature does not fit with the classical Kalman Filter tuning.
Figure \ref{fig:est_problem} shows a graphical representation of the proposed approach. 

The novel approach proposed in this work represents a leap forward in the field, and represents a timely contribution, for two main reasons: (1) vehicle manufacturers already use complex multibody simulators for other purposes like simulation and design, thus no additional modeling activity is needed for devising estimation-oriented simplified models; complex model allows us to consider ``fine effects'' like vibrations due to the engine or non-ideal behaviour due to suspensions geometry; (2) effective processing units like GPUs, needed to run such simulators, start to appear also in vehicles already on the market for self-driving features and computer vision tools like parking cameras. 

\textbf{Challenges.} Along with these benefits, the proposed approach offers important challenges. (1) In order to exploit the fidelity of a vehicle digital twin, it should be properly tuned, but model-based approaches like Kalman Filter formulas are no longer available, as no simple parametric model is given. Therefore, a data-driven simulation-error-minimization approach is employed, in which a cost function based on ground truth measures is minimized employing a Bayesian optimization routine (see \cite{FRAZIER_2018} and \cite{SHAHRIARI_2015}), which is well-known to be very efficient when the computational burden of each function evaluation is the main limitation.  Bayesian optimization is not completely new in the filter tuning context, see ad example \cite{chen2019kalman}, in which it is employed to tune all noise covariance parameters from available data. (2) Given the presence of a vehicle simulator in the loop, the number of states and output variables grows very fast, creating a high-dimensional tuning problem. In our proposal, a linear correction term is employed, where a matrix maps the output innovation into a suitable correction of the state estimates. In addition, to reduce the problem dimension, sparsity is enforced by means of the a-priori knowledge about the most important vehicle dynamics relationship of interest, exploiting the most important results taken from the literature.


\textbf{Structure.} The remainder of the paper is as follows. Section \ref{sec:architecture} describes in detail the proposed observer architecture. Section \ref{sec:obs_tuning} discusses the data-driven approach employed to tune the observer parameters. In Section \ref{sec:exp}, the architecture is tuned and validated on a case study based on real data to highlight the feasibility of the proposed approach. Finally, the section is concluded discussing a quantitative comparison with a benchmark employing simplified vehicle models. The paper is ended with some concluding remarks.

\section{Twin-based observer: architecture}
\label{sec:architecture}
\begin{figure*}[h!]
    \centering
    \includegraphics[width=\textwidth]{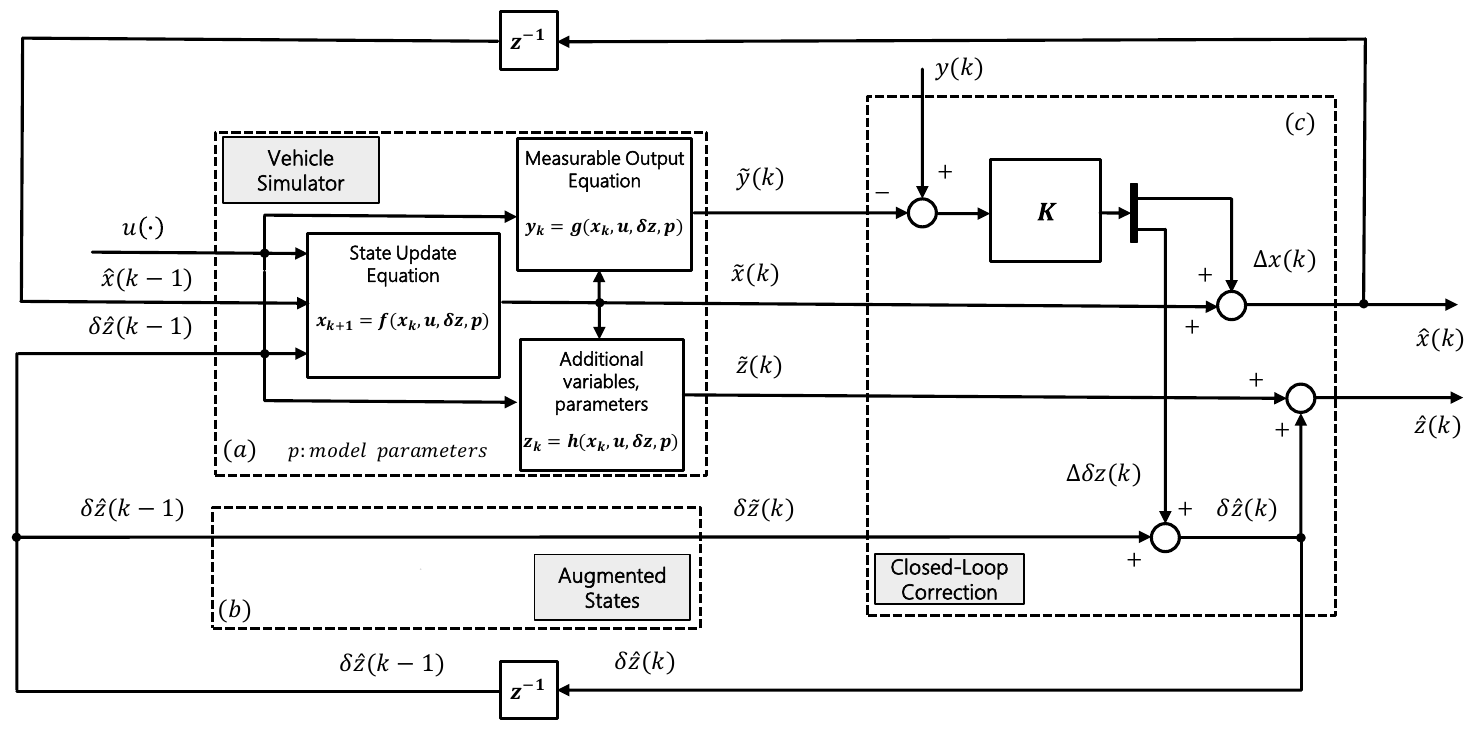}
    \caption{Twin-in-the-loop observer architecture: (a) vehicle simulator, which plays the role of the predictor; (b) external extended state vector including additional variables to estimate; (c) closed-loop correction block.}
    \label{fig:observer_architecture}
\end{figure*}

This section describes the proposed twin-based observer architecture, which is depicted in Fig. \ref{fig:observer_architecture}. 
Overall, three main modules can be identified, namely the vehicle simulator (a), an extended state to manage additional variables (b), and the closed-loop correction algorithm (c). All these three modules will be discussed in the next part of the section.
\subsection{Vehicle Simulator}
As anticipated before, the proposed solution focuses on the employment of a vehicle simulator as a predictor model inside the closed-loop observer framework. 
In Fig. \ref{fig:observer_architecture}, a simplified representation of the internal structure of the simulator is reported. The simulator block is fed with the external inputs $u(\cdot)$, the past corrected simulator state variables $\hat{x}(\cdot)$ and the corrected extended states  $\delta \hat{z}(\cdot)$, which are fed-back to the digital twin (see Section \ref{subsec:aug_state}).
The general equations of the vehicle simulator, in discrete time form with index $k$, are summarized as:
\begin{equation}
\begin{aligned}
   & \Tilde{x}(k)=f(\hat{x}(k-1),u(k-1),\delta\hat{z}(k-1),p)\\
    &\Tilde{y}(k)=g(\hat{x}(k),u(k),\delta\hat{z}(k-1),p)\\
    &\Tilde{z}(k)=h(\hat{x}(k),u(k),\delta\hat{z}(k-1),p)
\end{aligned}
\end{equation}
where $f$, $g$ and $h$ are black-box functions for us, while $y$, $z$ and $p$ are respectively measured output, additional estimated variables, and the whole set of parameters of the vehicle simulator.

Commonly, in the available commercial products, different inputs can be used to control the vehicle, in particular during braking and traction. The presence of such degrees of freedom allows the selection of the best solution depending on the type of information available on-board. As an example, the driver pedal request and the pressure applied at each corner are possible solutions to replicate the braking maneuver. 
Finally, it is important to stress that the usage of the vehicle simulator produces a huge increase of the model complexity, in particular concerning the number of state variables in the prediction model. This complexity increment should be properly handled in the closed-correction module design and in the corresponding tuning procedure, as explained respectively in Section \ref{subsec:cl_corr} and Section \ref{sec:obs_tuning}. 
\subsection{Augmented state}
\label{subsec:aug_state}
Estimation problems in automotive field usually concern standard state variables, like speeds and side-slip angles, which are labelled with $x$ in this work. This is not the only case: the estimation of vehicle parameters, e.g. vehicle mass and inertiae, and other non measurable signals, like tire-road contact forces, is of paramount importance for many control applications.
To cope with such problems we exploited a standard trick (see \cite{BAFFET20091255} for example), namely the addition of extended state variables, called $\delta z$, as shown in Fig. \ref{fig:observer_architecture}. These states lie outside the vehicle simulator and allow the real-time correction of both parameters and variables of interest based on the output innovation. We propose an additive solution: these degrees of freedom are used to correct the values $\tilde{z}$ provided by the simulator outputs, which can not be updated directly inside the digital twin, as done with the built-in state variables $x$. The corrected values are computed as:
\begin{equation}
    \hat{z}(k) = \tilde{z}(k) + \delta \hat{z}(k).
\end{equation}
These being artificial variables introduced for estimation purposes, they are associated to fictitious constant dynamic equations:
\begin{equation}
    \delta \tilde{z}(k+1)=\delta \hat{z}(k).
   \label{eq:augmented state equation}
\end{equation}
Overall, this structure results in the augmented state vector, defined as a column vector,
$
    x^{aug}=
    \begin{pmatrix}
    x^T, \ 
    \delta z^T
    \end{pmatrix}^T
$
, which is corrected exploiting the innovation, as explained in the next section. Finally, it is fundamental that the corrected states are introduced inside the vehicle simulator, so as to affect in turn the prediction phase, which is possible using ad-hoc simulator inputs.
\subsection{Closed-loop correction}
\label{subsec:cl_corr}
A Kalman Filter inspired matrix $K\in\mathbb{R}^{n\times p}$  is used to map the innovation, namely the difference between measured and predicted output variables $y-\Tilde{y}$, to state corrections $\Delta x^{aug} = \left( \begin{smallmatrix} \Delta x^T,\,\Delta \delta z^T
\end{smallmatrix}\right)^T$, through the equation
\begin{equation}
   \Delta x^{aug}(k)=K(y(k)-\Tilde{y}(k)).
\end{equation}
Then, as depicted in Fig. \ref{fig:observer_architecture}, the corrected augmented state vector $\hat{x}^{aug}$ is obtained from the predicted one $\Tilde{x}^{aug}$ as
\begin{equation}
    \hat{x}^{aug}(k)=\Tilde{x}^{aug}(k)+\Delta x^{aug}(k).
\end{equation}
The main motivation behind the choice of a linear correction term is dealing with computational complexity, a problem that arises due to the need of a data-driven solution within a problem with high dimension of state and output vectors.
\section{Observer tuning}
\label{sec:obs_tuning}
This section describes the second innovative contribution, namely a data-driven methodology to tune the closed-loop matrix introduced in Section \ref{subsec:cl_corr}.
In classical model-based estimators, the knowledge of system equations, e.g. matrices in the linear case, may allow the computation of the closed-loop term, for example via riccati equations in the Kalman Filter formulation.
Given the unavailability of a simple mathematical model of the vehicle, a data-driven approach is here proposed: an offline iterative procedure optimizes the correction term by minimizing the estimation error on a batch of recorded data on the real vehicle.

Given the setup of Fig. \ref{fig:observer_architecture}, it is impossible to decouple the matrix to be optimized from the vehicle simulator, so that we cannot compute inputs and outputs about the former independently from the matrix itself. Given this limitation, we cast the problem as a \textit{simulation-error-minimization} (SEM) problem, in which the solution is found by iterating multiple simulations over a certain time horizon. 

To limit the computational load, we enforce a level of sparsity in the correction matrix. Given the peculiarity of the estimation problem, we propose a sparsity-enforcement criteria based on the a-priori knowledge about the system dynamics. The choice depends from the available sensor layout of the vehicle, meaning that a unique selection cannot be done at this stage.
Once that sparsity is enforced in the matrix $K\in\mathbb{R}^{n\times p}$ , we can look at the optimization variables as the vector $\tilde{k}\in\mathbb{R}^{\Tilde{n}}$, where $\tilde{n}$ is the number of free parameters. The core of the optimization problem is the definition of the cost function $J( \tilde{k})$: we propose a classical solution, namely the weighted sum of the 2-norm of the estimation error of a sub-set of the augmented state vector previously defined, in order allow a proper customization of the cost. The choice of the variables weighted in the cost function is also dictated by the sensors availability in the training data. The defined optimization problem is formalized as follow, where $n_x$ and $n_z$ are respectively the number of state simulator variables and extended ones accounted in the optimization:
\begin{multline}
\min_{\tilde{k}}J(\tilde{k})=
\\ \min_{\tilde{k}}\sum_{m=1}^{n_x}  w_m^x\norm{\zeta^x_m-\hat{\zeta^x_m}(\tilde{k})}_2 +\sum_{l=1}^{n_z}  w_l^z\norm{\zeta^z_l-\hat{\zeta^z_l}(\tilde{k})}_2 .
\label{eq: cost function opt}    
\end{multline}
In \eqref{eq: cost function opt}, $\zeta^x_m$ and $\hat{\zeta^x}$ represent respectively the ground truth and estimated simulator state variables accounted in the training phase. Likewise, the extended states are labelled respectively with  $\zeta^z_l$ and $\hat{\zeta^z_l}$.
Finally, $w=\begin{smallmatrix}(w_1^x &\ldots & w_{n_x}^x,w_1^z&&\ldots&w_{n_z}^z)\end{smallmatrix}\in \mathbb{R}^{n_x+n_z}$ is the vector containing all the weighting coefficients.

In the SEM approach previously defined, the evaluation of the cost function \eqref{eq: cost function opt} requires the simulation of the whole estimation environment on the training dataset. Such computational burden should be efficiently faced by the numerical solution. 
Indeed, classical gradient-based non-convex optimization techniques present two shortcomings: (a) the unavailability of an analytical cost function formulation with respect to the decision variables requires the estimation of first and second order derivatives through additional cost function evaluations; (b) they easily fall into local minima without a sufficient exploration of the decision space.
Motivation (a), considering that the number of cost function evaluations is a scarce resource, represents the most critical issue for the application of such algorithms. Thus, we propose the employment of a gradient-free global optimization strategy inspired to \textit{Bayesian optimization} \cite{SHAHRIARI_2015,FRAZIER_2018}. In brief, the main idea of Bayesian optimization is to iterate between a learning and an optimization stage. During learning, a \textit{surrogate model}, typically a Gaussian process (GP),  approximating the objective function $J$ is estimated using all the available evaluations of $J$. During the optimization phase instead, the surrogate of the objective function is used to select the next most interesting point to evaluate. This is achieved by optimizing the so-called \textit{acquisition function} $\mathcal{A}$, constructed based on the surrogate model. Here, among the possible acquisition functions, we select the \textit{Expected Improvement} \cite{FRAZIER_2018}. This optimization algorithm is well-known to be efficient, providing fast convergence around the global optimal solution within a limited number of iterations.

Algorithm \ref{alg:opt} reports the pseudo-code containing the main steps of the Bayesian optimization algorithm, as it is employed in our application.
\begin{algorithm}[h!]
\DontPrintSemicolon
\KwInput{initial states: x(0), $\delta$z(0) , weights: $w$, Bayesian Optimization hyper-parameters: $N_{iter}$, $N_{init}$}
\KwOutput{optimal observer tuning $\tilde{k}^*$}
\KwData{experimental data $\mathcal{D}=\left\{u,y,\zeta^x,\zeta^z\right\}$}
\textbf{Extract} $N_{init}$ uniformly distributed random configurations of observer parameters: $\mathcal{K}=\left\{\tilde{k}_1, \,\ldots\,,\tilde{k}_{N_{init}}\right\}$\\
\textbf{Simulate} the estimation scheme fed by $\mathcal{D}$ for any element of $\mathcal{K}$ and \textbf{evaluate} the cost function: $\mathcal{J}=\left[ J(\tilde{k}_1),\,\ldots\,,J(\tilde{k}_{N_{init}})\right]$\\ 
\textbf{Initialize} $i=N_{init}$\\
   \While{$i<N_{iter}$}
   {
     \textbf{Compute} a GP approximation of $J$ based on the dataset $\left\{ \mathcal{K},\mathcal{J}\right\}$\;
     \textbf{Define} the acquisition function $\mathcal{A}\left(\tilde{k}\lvert\left\{ \mathcal{K},\mathcal{J}\right\}\right)$ \;
     \textbf{Compute} the next point $\tilde{k}^{\rightarrow}$ as: $\tilde{k}^{\rightarrow} = arg\min_{\tilde{k}} \mathcal{A}\left(\tilde{k}\lvert\left\{ \mathcal{K},\mathcal{J}\right\}\right)$  \;
     \textbf{Simulate} the estimation scheme fed by $\mathcal{D}$  with $\tilde{k}^{\rightarrow}$ and \textbf{evaluate} the cost function $J(\tilde{k}^{\rightarrow})$\;
     \textbf{Enlarge} the dataset: $\mathcal{K} = \mathcal{K}\cup \{ \tilde{k}^{\rightarrow}\}$, $\mathcal{J} = \mathcal{J}\cup [J(\tilde{k}^{\rightarrow})]$\;
     \textbf{$i=i+1$}\;
   }
\textbf{Compute} the optimal solution as: $\tilde{k}^*=arg\min_{\mathcal{K}} \mathcal{J}$
\caption{Bayesian optimization for twin-based observer tuning}
\label{alg:opt}
\end{algorithm}
%

\section{Experimental Case Study}
This section shows the effectiveness of the combination between the proposed estimation architecture, described in Section \ref{sec:architecture}, and the data-driven tuning method discussed in Section \ref{sec:obs_tuning}, through an experimental case study. After a description of the employed experimental setup in Section \ref{subsec:exp_setup}, a complete training procedure will be discussed in Section \ref{subsec:training}. Finally, the estimation results are analysed in Section \ref{subsec:testing} through a comparison with a baseline solution employing a simplified model, defined in Section \ref{subsec:bench_def}.
\label{sec:exp}
\subsection{Experimental setup}
\label{subsec:exp_setup}
The employed vehicle is equipped with: 1) a six-axis Inertial Measurement Unit (IMU), 2) optical sensors for vehicle velocities, 3) four wheel encoders measuring wheel rotational speeds, and 4) four in-wheel force transducers, necessary to calibrate the algorithm for the force estimation problem. Moreover, in order to properly drive the vehicle simulator, additional physical and virtual sensors are available, so as to access steering wheel angle, brake pressure at each corner, engine torque and the gear engaged.

The experimental campaign has been conducted on a sport car in a handling circuit, performing multiple laps with different driver behaviours, from standard to more aggressive maneuvers. The estimation scheme is based on the \texttt{VI-CarRealTime} (\texttt{VI-CRT}) \cite{vi-grade} vehicle simulator, which has been provided and fine-tuned by the car-manufacturer partner of this work. For confidentiality reasons, we cannot further specify the model of the car and track.

The focus of this experimental case study concerns classical state variables, namely longitudinal vehicle speed $v_x$ and side-slip angle $\beta=\arctan{\left(\frac{v_y}{v_x}\right)}$, where $v_y$ is the lateral vehicle speed. Moreover, also tire-road contact forces in the x-y plane $F^{ij}_x$ and $F^{ij}_y$, where $i=f$ (front), $r$ (rear) and $j=l$ (left), $r$ (right), are dealt with. These latter variables are introduced through the extended state approach discussed in Section \ref{subsec:aug_state} because they are not state variables of the digital twin. Thus, the augmented state vector defined in Section \ref{subsec:aug_state} is $
    x^{aug}=
    \begin{pmatrix}
    x^T, \delta F^T
    \end{pmatrix}^T
    =
    \begin{pmatrix}
    x^T, \delta F^{fl}_x, \ldots 
    \delta F^{rr}_y\\
    \end{pmatrix}^T
$
where $\delta F$ contains all the extended states related to the eight tire road forces of interest.

\subsection{Observer tuning}
\label{subsec:training}
The employed vehicle simulator has 14 degrees of freedom, meaning that 28 states can be manipulated. Among this list, we are particularly interested in the longitudinal and lateral vehicle velocities $v_x$, $v_y$, the yaw-rate $\dot{\psi}$ and the wheel angular velocities $\omega^{ij}$. Given the 8 extended states previously introduced to account for tire road forces, the augmented state vector has overall 36 elements, namely $x^{aug}\in\mathbb{R}^{36}$. Among the sensors available in the layout described in the previous section, we selected a suitable subset to build the measured output vector as follow:
\begin{equation}
    y =
    \begin{pmatrix}
    a_x,\;
    a_y,\;
    \dot{\psi},\;
    \omega^{fl},\;
    \omega^{fr},\;
    \omega^{rl},\;
    \omega^{rr}
    \end{pmatrix}^T
\end{equation}
Overall $y\in\mathbb{R}^7$: planar accelerations $a_x$, $a_y$ and vehicle yaw-rate $\dot{\psi}$ have been selected since they are strictly linked to the planar dynamics of interest, while the four wheel velocities $\omega^{ij}$ allows us to implement a direct feedback on the four available simulator states. The vehicle velocity and tire force sensors cannot be included in this vector since their availability is limited to the tuning phase, aimed to calibrate the matrix in the optimization.
Indeed, the variables weighted in the optimization problem defined in \eqref{eq: cost function opt} are:
\begin{equation}
    \zeta^x =   
    \begin{pmatrix}
    v_x\\
    v_y\\
    \dot{\psi}\\
    \omega^{fl}\\
    \omega^{fr}\\
    \omega^{rl}\\
    \omega^{rr}
  \end{pmatrix}
     \hspace{1cm}
     \zeta^z =   
    \begin{pmatrix}
    F^{fl}_x\\
    F^{fr}_x\\
    F^{rl}_x\\
    F^{rr}_x\\
    F^{fl}_y\\
    F^{fr}_y\\
    F^{rl}_y\\
    F^{rr}_y
  \end{pmatrix}.
\end{equation}
Overall, the matrix employed in the closed-loop correction algorithm has dimension  $K\in\mathbb{R}^{36\times7}$. At this stage we applied the proposed sparsity enforcement approach to reduce the problem dimension. As explained before, the selection of the parameters to optimize has been done by means of the a-priori knowledge about the most important links in the correction matrix, based also on the prior knowledge about the problem. The list of the selected parameters is the following: 1) $k_{v_x\omega}$: maps the wheel speed estimation error onto a correction on the vehicle longitudinal speed (same parameter for all corners); 2) $ k_{\dot{\psi}\dot{\psi}}$: maps the yaw-rate estimation error onto the corresponding state variable; 3) $k_{\omega\omega}$: maps the wheel speed estimation error onto the corresponding state variable (same parameter for all corners); 4) $k_{F_x a_x}$: maps the longitudinal acceleration estimation error onto a correction on the longitudinal force (same parameter for all corners); and 5) $k_{F_y \dot{\psi}}$: maps the yaw-rate estimation error onto a correction on the lateral forces (same parameter for all corners). Note that for this last parameter the sign between front and rear wheels has been reversed to exploit the a-priori knowledge about the effect on the rotational dynamics. Summarizing, the structure of the correction matrix is the following:
\begin{equation}
K =
    \begin{bmatrix}
     \sevenzeros\\
     \vdots & \vdots & \vdots & \vdots & \vdots & \vdots & \vdots\\
     \sevenzeros\\
     0 & 0 & 0 & k_{v_x\omega} & k_{v_x\omega} & k_{v_x\omega} & k_{v_x\omega}\\
     \sevenzeros\\
     \twozeros & k_{\dot{\psi}\dot{\psi}} & \fourzeros\\
     \threezeros & k_{\omega\omega} & \threezeros\\
     \fourzeros & k_{\omega\omega} & \twozeros\\
     \fivezeros & k_{\omega\omega} & 0\\
     \sixzeros & k_{\omega\omega}\\
     k_{F_x a_x} & \sixzeros \\
     k_{F_x a_x} & \sixzeros \\
     k_{F_x a_x} & \sixzeros \\
     k_{F_x a_x} & \sixzeros \\
     \twozeros & k_{F_y \dot{\psi}} & \fourzeros \\
     \twozeros & k_{F_y \dot{\psi}} & \fourzeros \\
     \twozeros & -k_{F_y \dot{\psi}} & \fourzeros \\
     \twozeros & -k_{F_y \dot{\psi}} & \fourzeros \\
    \end{bmatrix}
\end{equation}
where the first rows of zeros are related to all the state variables not considered, while the last fifteen are linked to $\zeta$ variables accounted in the optimization. 
From the whole matrix $K$ we can create the vector of free parameters to be optimized:
\begin{equation}
    \tilde{k}=\begin{pmatrix}
    k_{v_x\omega} & k_{\dot{\psi}\dot{\psi}} & k_{\omega\omega} & k_{F_x a_x} & k_{F_y \dot{\psi}}
    \end{pmatrix}.
\end{equation}
\begin{figure}[h!]
    \centering
    \includegraphics[width=1\columnwidth]{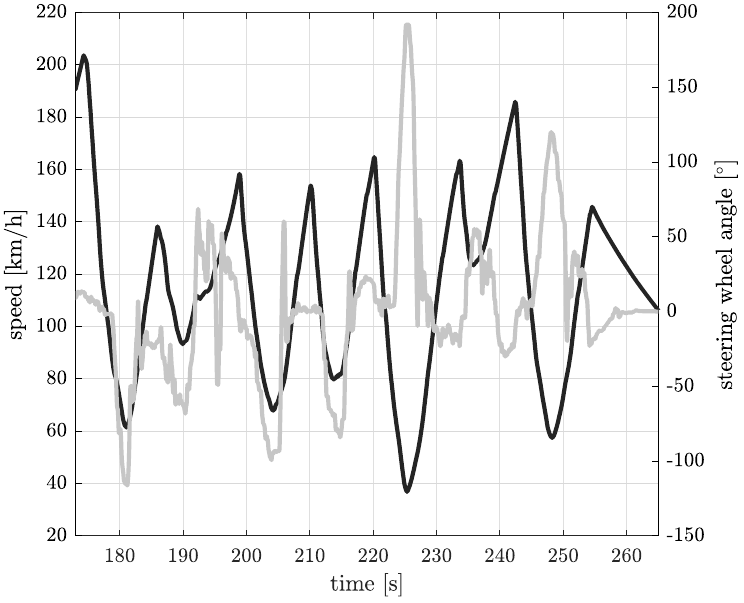}
    \caption{Speed (solid black line) and steering wheel angle (solid light grey line) profiles of the training dataset.}
    \label{fig:training_speed_profile}
\end{figure}
For each parameter of $\tilde{k}$, suitable bounds have been selected based on the a-priori information about magnitude between mapped output and state variables. As an example, for any parameters mapping the same variable, like $k_{\omega\omega}$, it is straightforward to select the lower and upper bounds as $0$ and $1$, respectively. The chosen bounds, used in the optimization routine, are reported in Table \ref{tab:bounds}. Moreover, the weights of the cost function have been chosen to make each term equally important. To this aim, the $2-norm$ of the measured signals is employed as follow:
\begin{table}[]
    \centering
    \begin{tabular}{ccc} 
    \toprule
        Parameter &  Lower Bound & Upper Bound \\
    \midrule
        $k_{\omega v_x}$ &  0 & 1 \\
        $k_{\dot{\psi} \dot{\psi}}$ &  0 & 1 \\
        $k_{\omega \omega}$ &  0 & 1 \\
        $k_{a_x F_x}$ & 0 &  100\\
        $k_{\dot{\psi} F_y}$ &  -100 & 0\\
    \bottomrule
    \end{tabular}
    \caption{Optimized closed-loop matrix parameters bounds.}
    \label{tab:bounds}
\end{table}
\begin{equation}
 w_i=\frac{1}{\norm{\zeta_i}_2}, \hspace{1cm }\forall i=1,\ldots,n_x+n_z.
 \label{eq:weights}
\end{equation}

For the tuning phase, a single lap has been taken, whose time history is shown in Fig. \ref{fig:training_speed_profile}. As it can be noticed from both speed and steering profiles, it represents an aggressive driving style, thus an informative dataset.
\\
Finally, Algorithm \ref{alg:opt} has been run to obtain the optimal matrix parameters, with the following Bayesian parameters:
\begin{equation}
    N_{iter}=100 \hspace{1cm} N_{init}=4.
\end{equation}
The achieved optimal parameters are summarised in Table \ref{tab:optimal}.
\begin{table}[]
    \centering
    \begin{tabular}{cc} 
    \toprule
        Parameter &  Value \\
    \midrule
        $k_{\omega v_x}$ &  $0.0808$ \\
        $k_{\dot{\psi} \dot{\psi}}$ &  $0.1328$ \\
        $k_{\omega \omega}$ &  $0.9593$ \\
        $k_{a_x F_x}$ & $98.42$ \\
        $k_{\dot{\psi} F_y}$ &  $-75.32$ \\
    \bottomrule
    \end{tabular}
    \caption{Optimal closed-Loop matrix parameters.}
    \label{tab:optimal}
\end{table}

\begin{figure*}[h!]
\centering
\begin{subfigure}[b]{0.49\textwidth}
    \centering
     \includegraphics[width=\textwidth]{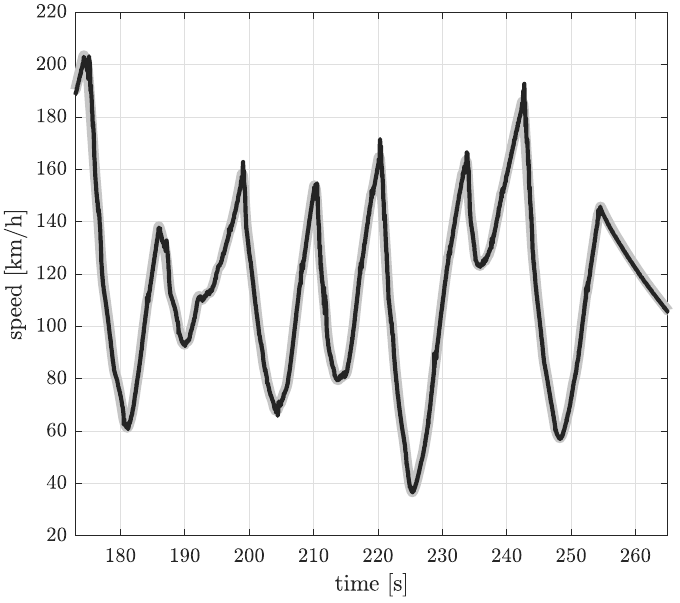}
    \caption{Longitudinal speed.}
    \label{fig:training_vx}
\end{subfigure}
\hfill
\begin{subfigure}[b]{0.49\textwidth}
    \centering
     \includegraphics[width=\textwidth]{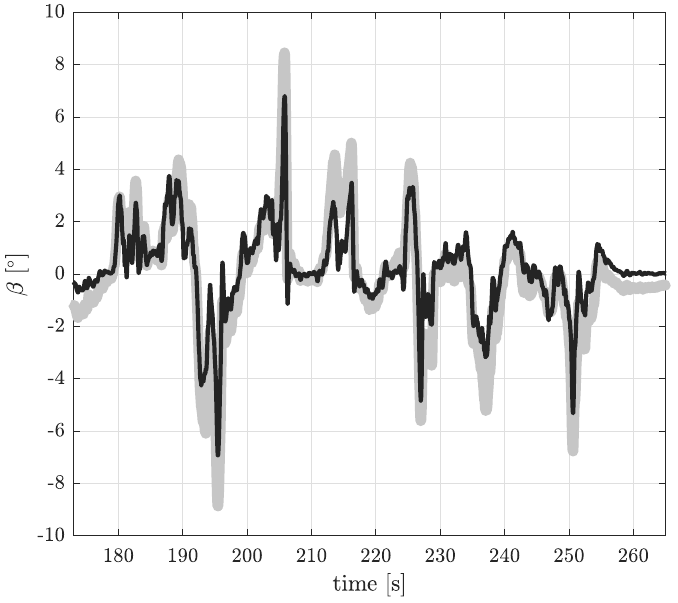}
    \caption{Vehicle side-slip angle.}
    \label{fig:training_beta}
\end{subfigure}
\begin{subfigure}[b]{0.49\textwidth}
    \centering
     \includegraphics[width=\textwidth]{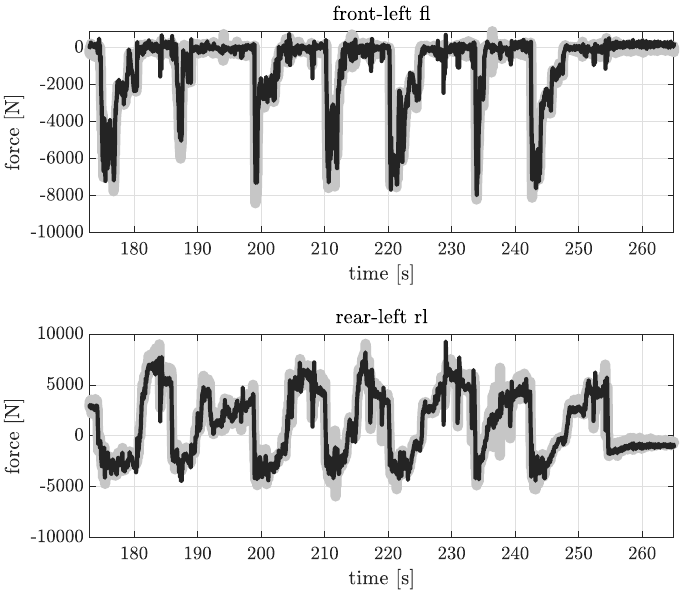}
    \caption{Longitudinal forces (front-left on top and rear-left on bottom).}
    \label{fig:training_fx}
\end{subfigure}
\hfill
\begin{subfigure}[b]{0.49\textwidth}
    \centering
     \includegraphics[width=\textwidth]{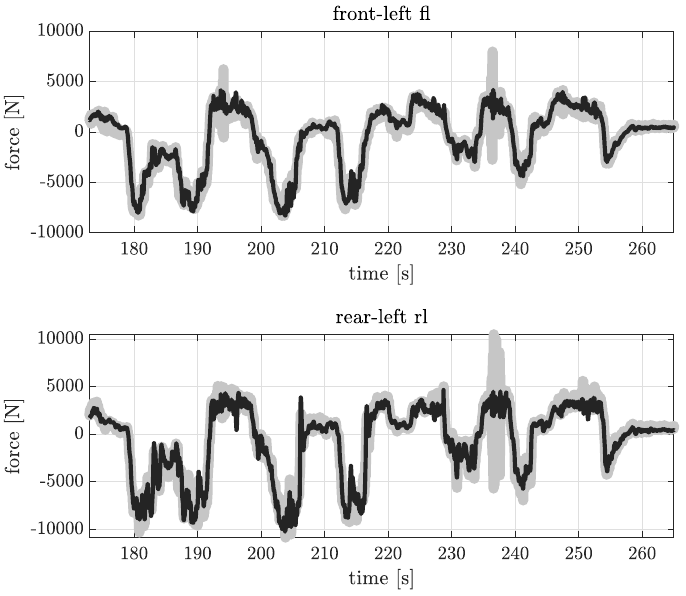}
    \caption{Lateral forces (front-left on top and rear-left on bottom).}
    \label{fig:training_fy}
\end{subfigure}
    \caption{Estimation performance on a training dataset: longitudinal speed (a), vehicle side-slip angle (b), longitudinal forces (c) and lateral ones (d). Comparison between experimental data  (solid light grey line) and twin-based estimate  (solid black line).}
    \label{fig:training_perf}
\end{figure*}

Fig. \ref{fig:training_perf} shows the time histories of the obtained optimal tuning on the training dataset: Figures \ref{fig:training_vx} and \ref{fig:training_beta} report longitudinal speed and side-slip angle, while Figures \ref{fig:training_fx} and \ref{fig:training_fy} display the estimation results about longitudinal and lateral forces. For saving space, only the left wheels are shown in force-related graphs. 
\subsection{Benchmark definition}
\label{subsec:bench_def}
\begin{figure*}[h!]
\centering
\begin{subfigure}[b]{0.49\textwidth}
    \centering
     \includegraphics[width=.7\textwidth]{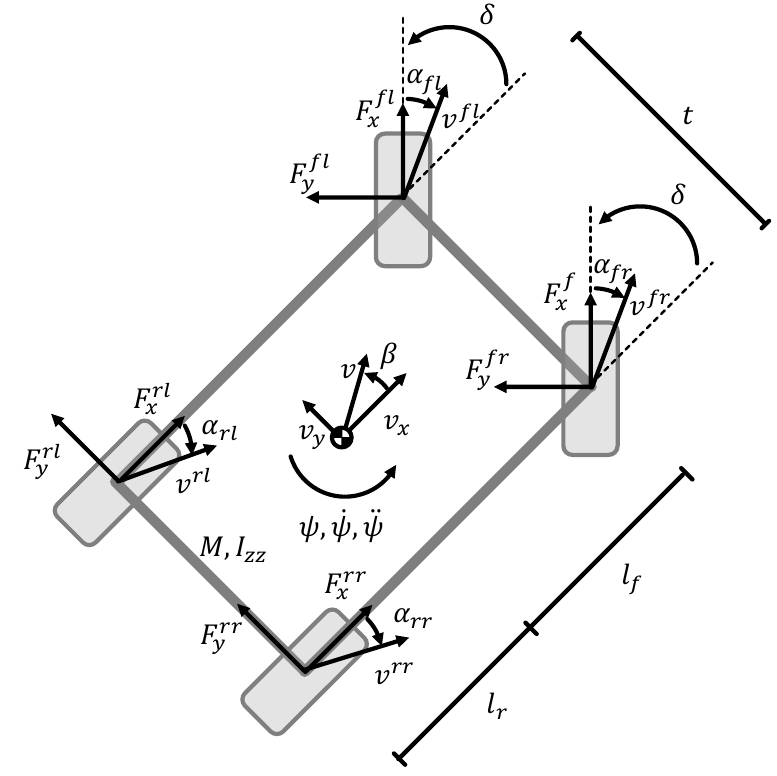}
    \caption{Double-track vehicle planar model.}
    \label{fig:bench_double_track}
\end{subfigure}
\hfill
\begin{subfigure}[b]{0.49\textwidth}
    \centering
     \includegraphics[width=.8\textwidth]{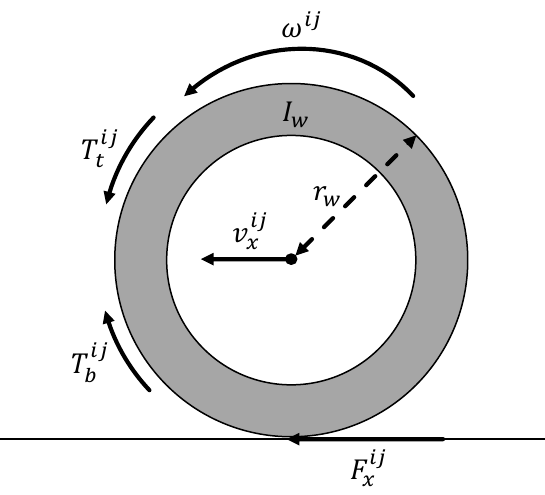}
    \caption{Wheel dynamics model.}
    \label{fig:bench_wheel_model}
\end{subfigure}
\caption{Schematics of the employed models inside the benchmark: (a) double-track planar model; (b) wheel dynamics model.}
\label{fig:benchmark}
\end{figure*}

In order to make a fair evaluation of the approach, this section describes the model-based benchmark employed in the comparison of Section \ref{subsec:testing}. We keep the same estimation architecture and tuning method, substituting the vehicle simulator with a simplified vehicle model, which has been constructed based on the typical models employed in literature. It combines a planar double-track model, used for example in \cite{DOUMIATI_ACC2010}, and four wheel dynamics models \cite{rezaeian2014novel}. A graphical representation of both subsystems is shown in Fig. \ref{fig:benchmark}. We remark that the employed physical parameters are obtained both from a-priori information about the vehicle and proper identification experiments on real-data, e.g. for the simplified Pacejka models of \eqref{eq:pacejka model}.

Starting from the double-track model of Fig. \ref{fig:bench_double_track}, a third order dynamical system with state variables $x=\lvert v_x,\;v_y,\;\dot{\psi}\lvert^T$ can be derived as
\begin{equation}
    \begin{aligned}
       &\dot{v}_x=\frac{F^T_x}{M}+v_y\dot{\psi}\\
       &\dot{v}_y=\frac{F^T_y}{M}-v_x\dot{\psi}\\
       &\ddot{\psi}=\frac{M^T_z}{I_{zz}}.\\
    \end{aligned}
    \label{eq:double track model}
\end{equation}
Inside \eqref{eq:double track model}, $M$ and $I_{zz}$ represent respectively the overall vehicle mass and the vehicle moment of inertia around the vertical axis, while $F^T_x$, $F^T_y$, and $M^T_z$ are computed according to the following expressions:
\begin{equation}
    \begin{aligned}
       & \begin{aligned}F^T_x=&(F^{fl}_x+F^{fr}_x)\cos(\delta)-(F^{fl}_y+F^{fr}_y)\sin(\delta)\\&+F^{rl}_x+F^{rr}_x \end{aligned}\\
       & \begin{aligned}F^T_y=&(F^{fl}_x+F^{fr}_x)\sin(\delta)+(F^{fl}_y+F^{fr}_y)\cos(\delta)\\&+F^{rl}_y+F^{rr}_y \end{aligned}\\
       &\begin{aligned}
               M^T_z=l_f(F^{fl}_y+F^{fr}_y)\cos(\delta)+\frac{t}{2}(F^{fl}_y-F^{fr}_y)\sin(\delta)\\-\frac{t}{2}(F^{fl}_x-F^{fr}_x)\cos(\delta)
       +l_f(F^{fl}_x+F^{fr}_x)\sin(\delta)\\-l_r(F^{rl}_y+F^{rr}_y)-\frac{t}{2}(F^{rl}_x-F^{rr}_x) , 
       \end{aligned}
    \end{aligned}
    \label{eq:chassis forces and moments}
\end{equation}
where $\delta$ is an input of the model, namely the steering angle at the front wheels, considered equal on both corners for simplicity.
Moreover, $l_f$ and $l_r$ represent the distance of the centre of mass from the front  and rear axles respectively, while $t$ is the average vehicle track width. All the longitudinal and lateral forces are worked out employing the simplified Pacejka magic formula in \eqref{eq:pacejka model}, accounting only the pure longitudinal and lateral slip conditions.
\begin{equation}
    \begin{aligned}
       &
       \begin{aligned}
       F^{ij}_x = &f_x(\lambda^{ij},F^{ij}_z) = F^{ij}_z D_x\sin(C_x \arctan(B_x \lambda^{ij}-\\
       &-E_x(B_x \lambda^{ij}-\arctan(B_x \lambda^{ij}))))
       \end{aligned}\\\\
       &
       \begin{aligned}
       F^{ij}_y = &f_y(\alpha^{ij},F^{ij}_z)=F^{ij}_z D_y \sin(C_y \arctan(B_y \alpha^{ij}-\\
       &-E_y(B_y \alpha^{ij}-\arctan(B_y \alpha^{ij}))))
        \end{aligned}
    \end{aligned}
    \label{eq:pacejka model}
\end{equation}
In \eqref{eq:pacejka model}, $F^{ij}_z$ is the vertical load at each corner, computed with a simple load transfer model from vehicle longitudinal and lateral accelerations, as proposed also in \cite{viehweger2020vehicle}, and $\lambda^{ij}$ and $\alpha^{ij}$ are respectively the longitudinal and lateral slip at each wheel, computed as:
\begin{equation}
    \begin{aligned}
    & \lambda^{ij}= \frac{r^{ij}_w\omega^{ij}-v^{ij}_x}{\max(r^{ij}_w\omega^{ij},v^{ij}_x)}, 
    & \alpha^{ij} = \arctan{\left(\frac{v^{ij}_y}{v^{ij}_x}\right)}.
    \end{aligned}
    \label{eq:slip definition}
\end{equation}
In \eqref{eq:slip definition}, $v^{ij}_x$ and $v^{ij}_y$ are the components of the wheel centre velocity vector expressed in the wheel frame, while $r^{ij}_w$ and $\omega^{ij}$ are respectively the wheel radius and the wheel rotational velocity, which dynamics is described in \eqref{eq:wheel dyn}.
\\
Turning to the wheel subsystem of Fig. \ref{fig:bench_wheel_model}, the dynamics can be simply expressed as
\begin{equation}
    \dot{\omega}^{ij}=\frac{T^{ij}_t-T^{ij}_b-r^{ij}_wF^{ij}_x}{I_w},
    \label{eq:wheel dyn}
\end{equation}
where $F_x^{ij}$ is the longitudinal wheel force, as previously specified in \eqref{eq:pacejka model}. Finally, $T^{ij}_b$ and $T^{ij}_t$ represent respectively braking and traction torques applied at each wheel, defined as:
\begin{equation}
    \begin{aligned}
    &T^{ij}_b=k_b p^{ij}_b\\
    &T^{ij}_t=
    \begin{cases}
    0 \hspace{1.3cm} i=f,\,\,j = l,r\\
    \frac{T_m\Gamma(\sigma)}{2} \hspace{0.5cm}  i=r,\,\,j = l,r.
    \end{cases}
    \end{aligned}
    \label{eq:torques}
\end{equation}
In \eqref{eq:torques}, $p^{ij}_b$ is the braking pressure at each corner, an input variable, which is transformed into a torque via the coefficient $k_b$. 
Moreover, $T_m$ and $\sigma$ are other two inputs of the model, namely the produced engine torque and the actual engaged gear, the latter employed to compute the gear ratio $\Gamma$. Note that, a simple traction torque split between the real wheels has been employed, without considering a more realistic differential model. 

Overall, from \eqref{eq:double track model} to \eqref{eq:torques}, the system can be expressed as the following discrete-time state-space model:
\begin{equation}
\begin{aligned}
    &x^{bench}(k+1) = f_{bench}(x^{bench}(k),u^{bench}(k))\\
    &y^{bench}(k) = g_{bench}(x^{bench}(k),u^{bench}(k)) \\
\end{aligned}
\label{eq:state-space benchmark}
\end{equation}
where $f_{bench}$, $g_{bench}$ are suitable functions, and:
\begin{equation}
    x^{bench}=
    \begin{pmatrix}
    v_x\\
    v_y\\
    \dot{\psi}\\
    \omega^{fl}\\
    \omega^{fr}\\
    \omega^{rl}\\
    \omega^{rr}
    \end{pmatrix}
    ,\hspace{1cm} y^{bench} =
    \begin{pmatrix}
    a_x\\
    a_y\\
    \dot{\psi}\\
    \omega^{fl}\\
    \omega^{fr}\\
    \omega^{rl}\\
    \omega^{rr}
    \end{pmatrix}.
    \label{eq:state-space benchmark variables}
\end{equation}
Also for the benchmark, the estimation objective concerns both classical state variables, namely $v_x$ and $\beta$, and tire road forces, which are handled with the extended state approach explained before for the twin-based estimator. Moreover, the same sparsity is enforced in the correction matrix, resulting in the following structure $K \in \mathbb{R}^{15\times 7}$,
\begin{equation}
K =
    \begin{bmatrix}
     0 & 0 & 0 & k_{v_x\omega} & k_{v_x\omega} & k_{v_x\omega} & k_{v_x\omega}\\
     \sevenzeros\\
     \twozeros & k_{\dot{\psi}\dot{\psi}} & \fourzeros\\
     \threezeros & k_{\omega\omega} & \threezeros\\
     \fourzeros & k_{\omega\omega} & \twozeros\\
     \fivezeros & k_{\omega\omega} & 0\\
     \sixzeros & k_{\omega\omega}\\
     k_{F_x a_x} & \sixzeros \\
     k_{F_x a_x} & \sixzeros \\
     k_{F_x a_x} & \sixzeros \\
     k_{F_x a_x} & \sixzeros \\
     \twozeros & k_{F_y \dot{\psi}} & \fourzeros \\
     \twozeros & k_{F_y \dot{\psi}} & \fourzeros \\
     \twozeros & -k_{F_y \dot{\psi}} & \fourzeros \\
     \twozeros & -k_{F_y \dot{\psi}} & \fourzeros \\
    \end{bmatrix}
\end{equation}
with the same vector of optimized variables:
\begin{equation}
    \tilde{k}=\begin{pmatrix}
    k_{v_x\omega} & k_{\dot{\psi}\dot{\psi}} & k_{\omega\omega} & k_{F_x a_x} & k_{F_y \dot{\psi}}
    \end{pmatrix}.
\end{equation}
Finally, we run Algorithm \ref{alg:opt} with the same Bayesian parameters, weights and bounds of Section \ref{subsec:training} to work out the optimal observer parameters. The next section will compare the results of the two approaches on a testing dataset.
\begin{figure*}[]
\centering
\begin{subfigure}[b]{0.49\textwidth}
    \centering
     \includegraphics[width=\textwidth]{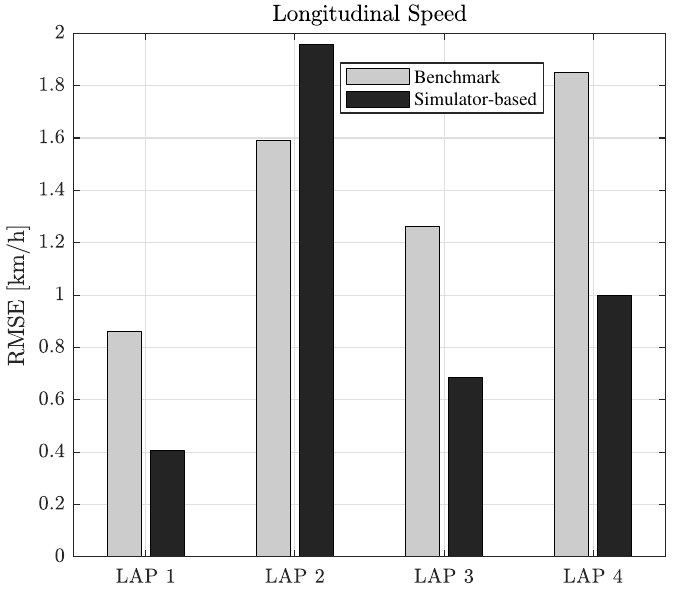}
    \caption{Longitudinal speed.}
    \label{fig:testing_rmse_vx}
\end{subfigure}
\hfill
\begin{subfigure}[b]{0.49\textwidth}
    \centering
     \includegraphics[width=\textwidth]{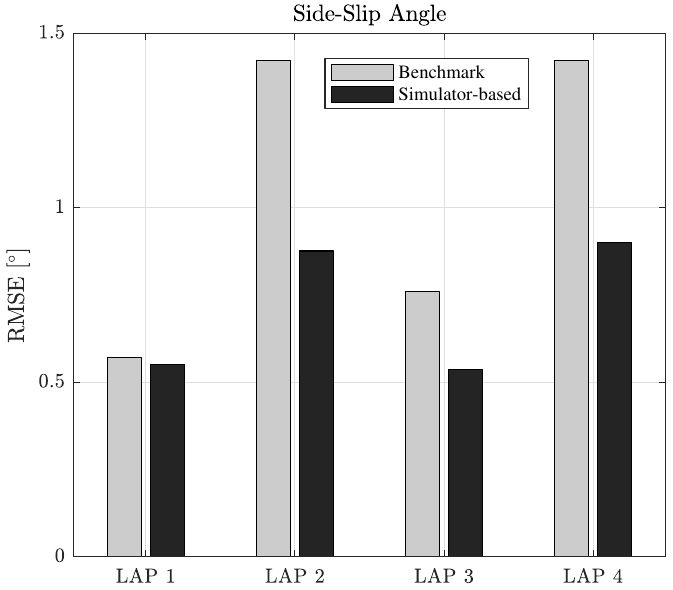}
    \caption{Vehicle side-slip angle.}
    \label{fig:testing_rmse_beta}
\end{subfigure}
\begin{subfigure}[b]{0.49\textwidth}
    \centering
     \includegraphics[width=\textwidth]{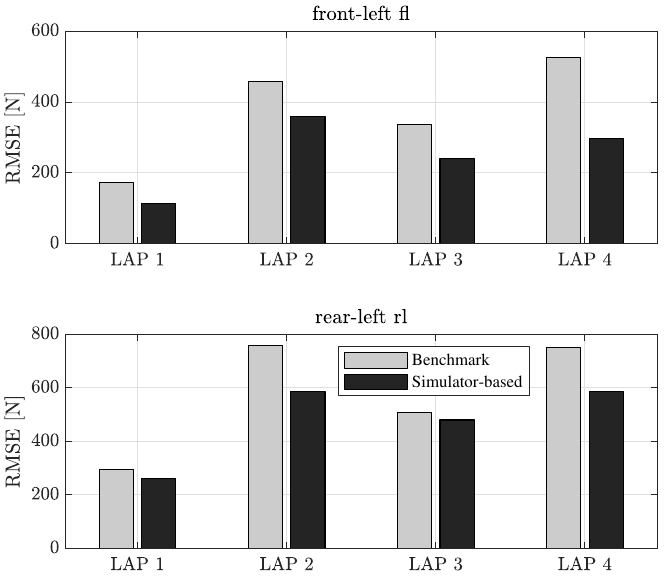}
    \caption{Longitudinal forces (front-left on top and rear-left on bottom).}
    \label{fig:testing_rmse_fx}
\end{subfigure}
\hfill
\begin{subfigure}[b]{0.49\textwidth}
    \centering
     \includegraphics[width=\textwidth]{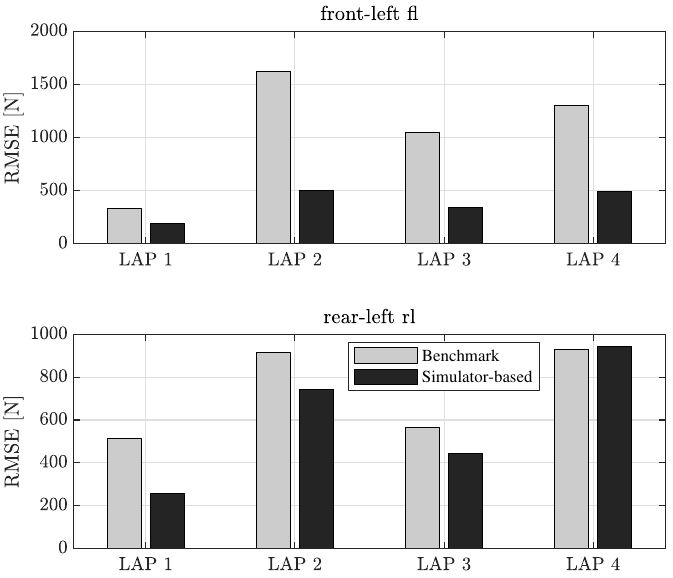}
    \caption{Lateral forces (front-left on top and rear-left on bottom).}
    \label{fig:testing_rmse_fy}
\end{subfigure}
    \caption{Root-Mean-Square-Error (RMSE) over a set of testing maneuvers. Comparison between twin-based (black bars) and benchmark (grey bars) schemes.}
    \label{fig:testing_rmse}
\end{figure*}
\subsection{Testing results}
\label{subsec:testing}
\begin{figure*}[]
\centering
\begin{subfigure}[b]{0.49\textwidth}
    \centering
     \includegraphics[width=\textwidth]{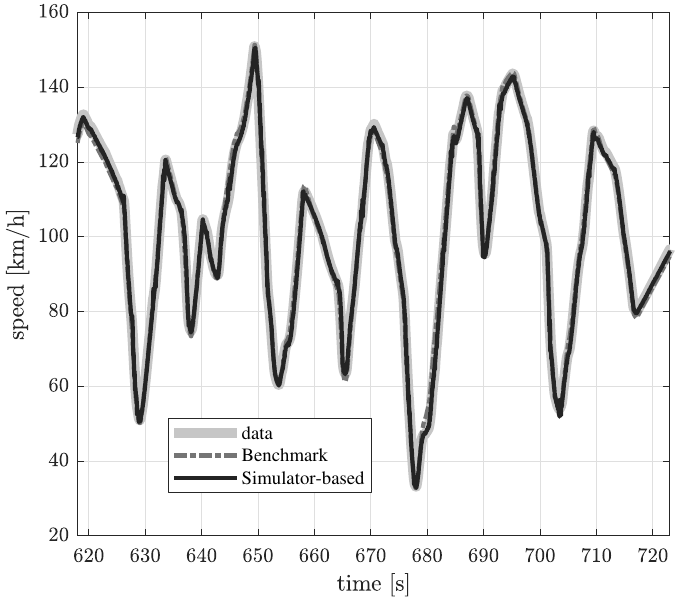}
    \caption{Longitudinal vehicle speed estimation on a whole lap.}
    \label{fig:testing_Vx}
\end{subfigure}
\hfill
\begin{subfigure}[b]{0.49\textwidth}
    \centering
     \includegraphics[width=\textwidth]{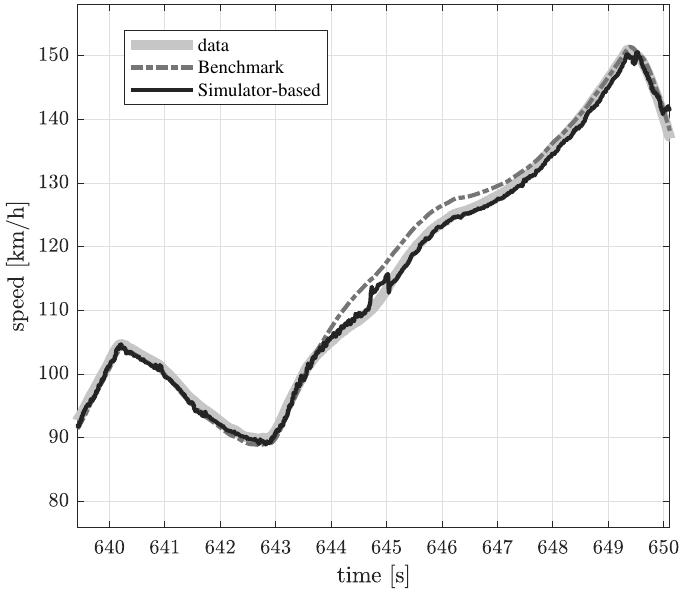}
    \caption{Longitudinal vehicle speed estimation on a zoomed section.}
    \label{fig:testing_Vx_zoom}
\end{subfigure}
\begin{subfigure}[b]{0.49\textwidth}
    \centering
     \includegraphics[width=\textwidth]{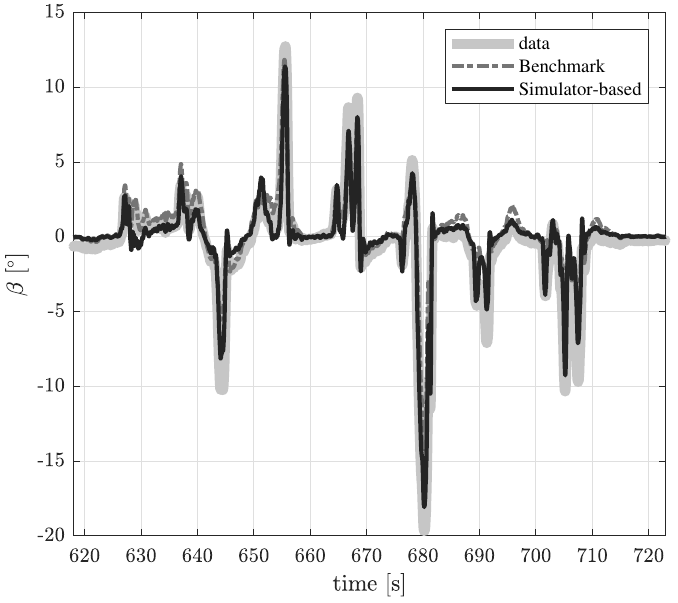}
    \caption{Vehicle side-slip angle estimation on a whole lap.}
    \label{fig:testing_Beta}
\end{subfigure}
\hfill
\begin{subfigure}[b]{0.49\textwidth}
    \centering
     \includegraphics[width=\textwidth]{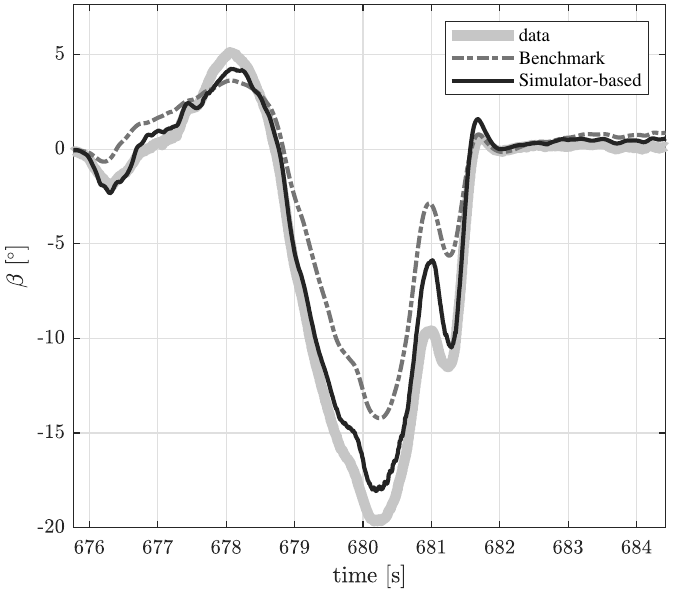}
    \caption{Vehicle side-slip angle estimation on a zoomed section.}
    \label{fig:testing_Beta_zoom}
\end{subfigure}
    \caption{Estimation performance on a testing dataset. Comparison among experimental data (solid light grey line), twin-based estimate (solid black line), and benchmark estimate (dash-dot dark grey line).}
    \label{fig:testing_Vx_Beta}
\end{figure*}

All the following results consider a set of four additional laps, used as testing dataset.
To have a general perspective, we look to the Root-Mean-Square-Error (RMSE) in each lap to compare the performance of the algorithms, as shown in Fig. \ref{fig:testing_rmse}. More precisely: Figs. \ref{fig:testing_rmse_vx} and \ref{fig:testing_rmse_beta} report longitudinal velocity and side-slip angle, while Figs. \ref{fig:testing_rmse_fx} and \ref{fig:testing_rmse_fy} highlight respectively longitudinal and lateral forces. Also in this case, only the two left corners are reported. As Fig. \ref{fig:testing_rmse} clearly shows, we can appreciate a significant performance improvement in the overall estimation, particularly concerning the front lateral forces in the more aggressive laps, namely $2$ and $4$. Nonetheless, there are situations in which the performance of the two algorithms are almost comparable or even slightly better for the benchmark one, e.g. looking to the longitudinal velocity estimation in the second lap or to the rear longitudinal forces in the third lap. However, we recall here that our purpose is not to \textit{always} guarantee the best performance, but to show an overall better behaviour, without the need of devising ad-hoc estimation oriented models.
\\
The analysis is enriched by illustrating the time histories of the estimation in the more aggressive lap, namely the fourth one. Fig. \ref{fig:testing_Vx_Beta} reports the estimation performance of longitudinal speed (Fig. \ref{fig:testing_Vx}) and side-slip angle (Fig. \ref{fig:testing_Beta}), with a highlight respectively in Figs. \ref{fig:testing_Vx_zoom} and \ref{fig:testing_Beta_zoom}.  From this graphs, we can notice the better performance achieved with the twin-based scheme for both estimated variables.

\begin{figure*}[]
\centering
\begin{subfigure}[b]{0.49\textwidth}
    \centering
     \includegraphics[width=\textwidth]{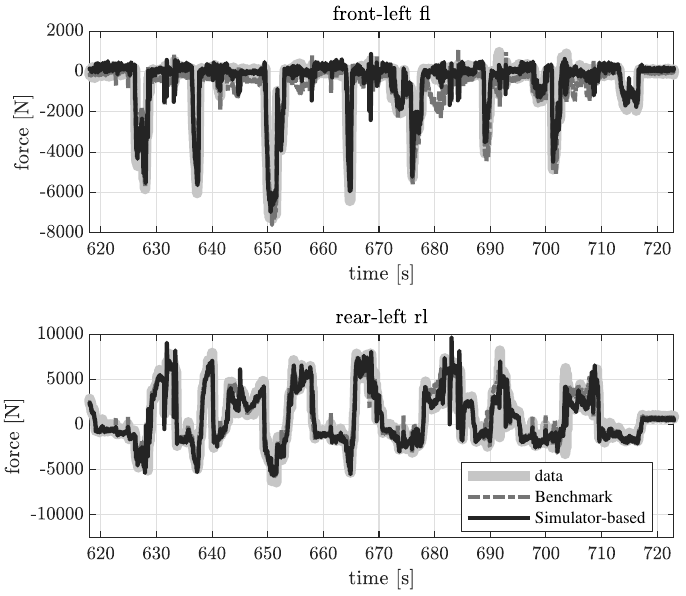}
    \caption{Longitudinal forces (front-left on top and rear-left on bottom) estimation on a whole lap.}
    \label{fig:testing_fx}
\end{subfigure}
\hfill
\begin{subfigure}[b]{0.49\textwidth}
    \centering
     \includegraphics[width=\textwidth]{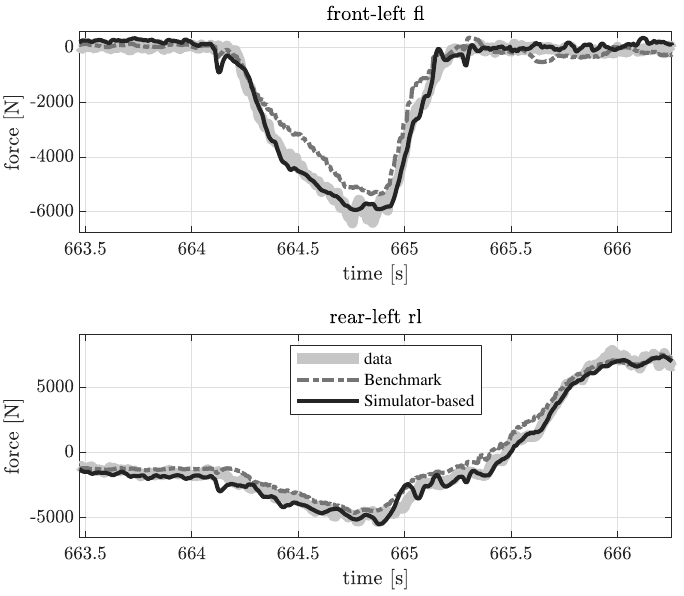}
    \caption{Longitudinal forces (front-left on top and rear-left on bottom) estimation on a zoomed section.}
    \label{fig:testing_fx_zoom}
\end{subfigure}
\begin{subfigure}[b]{0.49\textwidth}
    \centering
     \includegraphics[width=\textwidth]{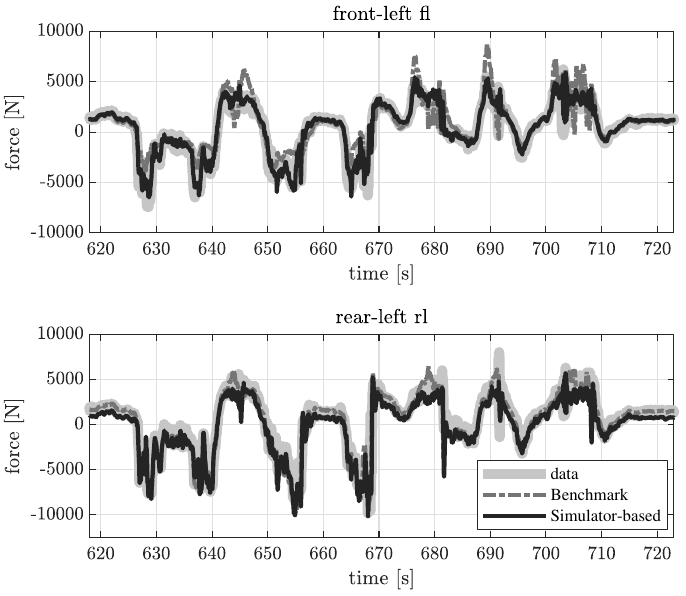}
    \caption{Lateral forces (front-left on top and rear-left on bottom) estimation on a whole lap.}
    \label{fig:testing_fy}
\end{subfigure}
\hfill
\begin{subfigure}[b]{0.49\textwidth}
    \centering
     \includegraphics[width=\textwidth]{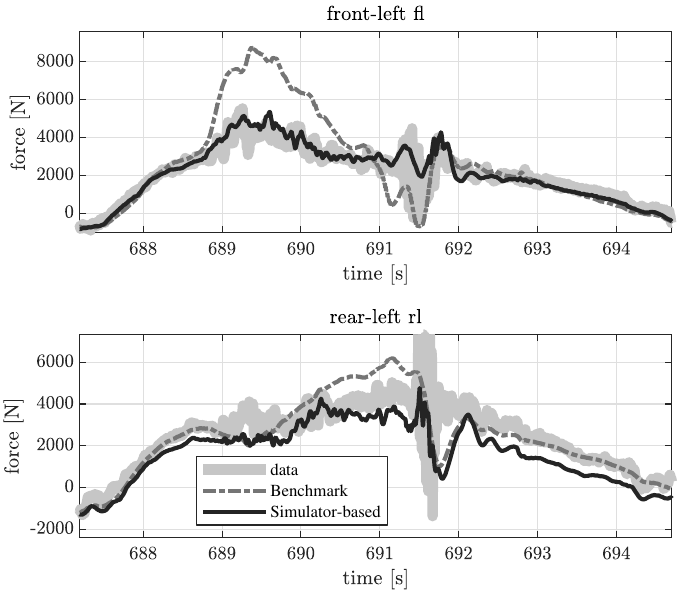}
    \caption{Lateral forces (front-left on top and rear-left on bottom) estimation on a zoomed section.}
    \label{fig:testing_fy_zoom}
\end{subfigure}
    \caption{Estimation performance on a testing dataset: comparison between twin-based approach and benchmark one about longitudinal forces [(a),(b)] and lateral ones [(c),(d)].  Comparison among experimental data  (solid light grey line), twin-based estimate  (solid black line), and benchmark estimate (dash-dot dark grey line).}
    \label{fig:testing_fx_fy}
\end{figure*}

Turning to the estimation of tire forces, Fig. \ref{fig:testing_fx_fy} shows the time series comparing the two algorithms, respectively in Figs. \ref{fig:testing_fx} and  \ref{fig:testing_fy} for longitudinal and lateral ones. From the highlights of Figs. \ref{fig:testing_fx_zoom} and \ref{fig:testing_fy_zoom}, we can appreciate also in this case the higher accuracy of the twin-based estimation algorithm.

\begin{figure*}[]
\centering
 \begin{subfigure}[b]{0.49\textwidth}
    \centering
    \includegraphics[width=\columnwidth,trim={2.5cm 10.5cm 2cm 9.8cm},clip]{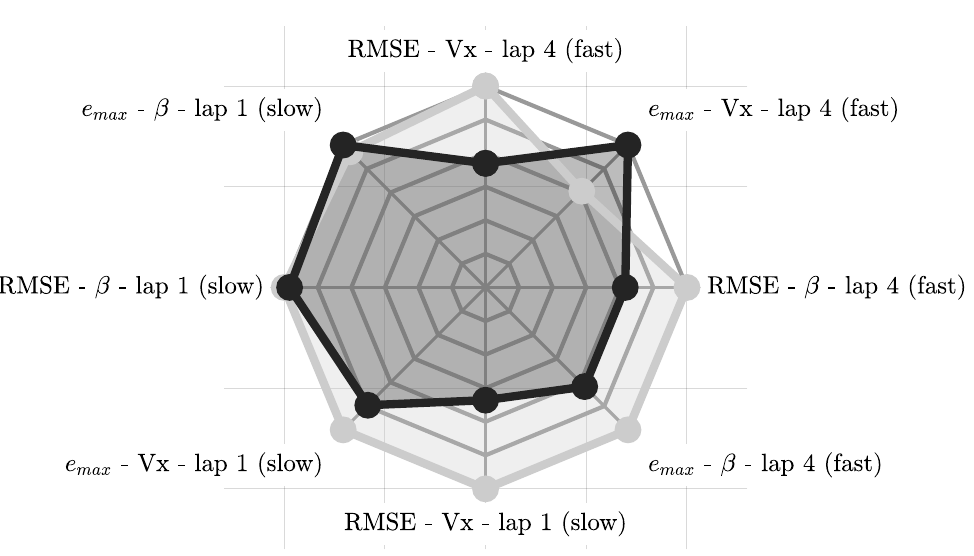}
    \caption{Longitudinal speed and vehicle side-slip angle.}
    \label{fig:spider_vx_beta}
\end{subfigure}
\begin{subfigure}[b]{0.49\textwidth}
    \includegraphics[width=\columnwidth,trim={2cm 10.5cm 1.8cm 9.8cm},clip]{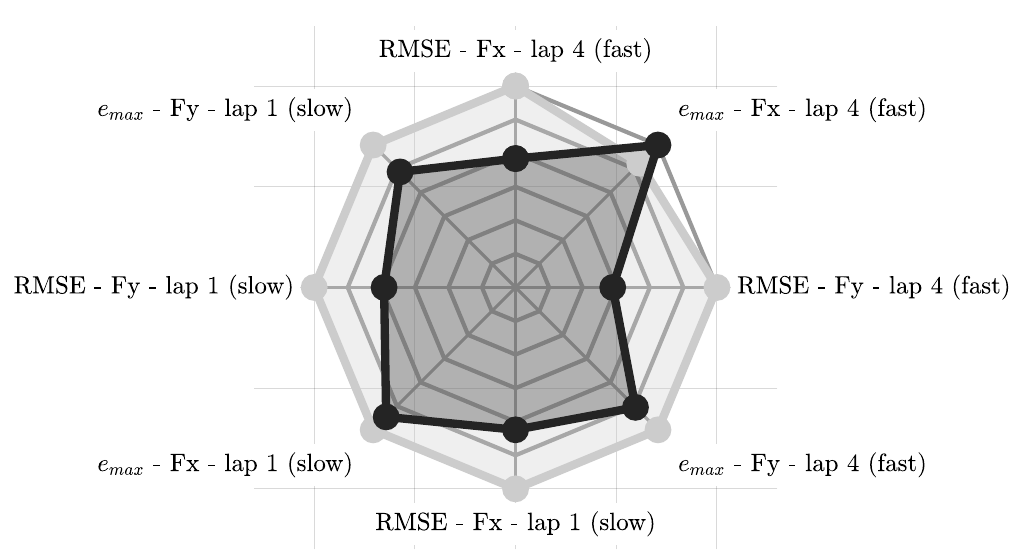}
    \caption{Longitudinal and lateral front-left forces.}
    \label{fig:spider_fx_fy}
\end{subfigure}
\caption{Spider plot comparing normalized estimation performance between twin-based approach (black graph) and benchmark one (grey graph). Root-Mean-Square-Error and maximum error are used as performance indexes in two different scenarios: a slow and a fast lap of the testing dataset.}
\label{fig:spider_plot}
\end{figure*}

To summarize the proposed results of the experimental case study, Fig. \ref{fig:spider_plot} shows the spider plot of the normalized estimation performance, looking at the RMSE and the maximum absolute estimation error in two different laps, namely a standard (lap 1) and a more aggressive one (lap 4). Fig. \ref{fig:spider_vx_beta} reports respectively longitudinal velocity and side-slip angle, while Fig. \ref{fig:spider_fx_fy} illustrates the longitudinal and lateral front-left forces. We can notice once more the effectiveness of the proposed twin-based approach, which is able to provide more precise estimations in almost all the considered performance indexes. It seems then that the additional value provided by the digital twin stands out when more aggressive maneuvers are performed by the driver, as there having an accurate description of the whole vehicle dynamics becomes particularly critical.

\section{Conclusions}
In this paper, a novel estimation framework of vehicle dynamics variables is proposed and tested. The architecture is that of classical closed-loop observers, where the role of the vehicle model is taken by a  multibody simulator, allowing to accurately estimate all the possible variables of interest. An extended state solution is also developed for embedding estimation of additional variables and parameters. We propose a linear structure for the correction block and a data-driven sparsity-inducing tuning algorithm, in order to keep the computational load under control.
An experimental case study is eventually discussed using a sport-car on a handling circuit. After showing the feasibility of the tuning pipeline, the results are compared with a benchmark solution employing a standard vehicle model. The results highlight the potential of the proposed approach, which tends to outperform the benchmark estimators, especially during aggressive manoeuvres as far as both kinematic planar variables or tire-road forces estimation are concerned.
Future work will be devoted to extensive experimental testing of the proposed approach on other vehicle dynamics estimation scenarios.

\bibliographystyle{elsarticle-harv}\biboptions{authoryear}
\bibliography{biblio}
\end{document}

%% file: new_comms.tex
\newcommand{\twozeros}{0 & 0}
\newcommand{\threezeros}{0 & 0 & 0}
\newcommand{\fourzeros}{0 & 0 & 0 & 0}
\newcommand{\fivezeros}{0 & 0 & 0 & 0 & 0}
\newcommand{\sixzeros}{0 & 0 & 0 & 0 & 0 & 0}
\newcommand{\sevenzeros}{0 & 0 & 0 & 0 & 0 & 0 & 0}
\newcommand{\norm}[1]{\left\lVert#1\right\rVert}
\SetKwInput{KwInput}{Input}                
\SetKwInput{KwOutput}{Output}              